\documentclass[preprint2, times]{aastex62}
\usepackage{float}
\usepackage{ulem}

\newcommand\msunyr{\rm {\it M}_{\odot}\,yr^{-1}}

\begin{document}

\title{Modeling the Protoplanetary Disks of Two Brown Dwarfs in the Taurus Molecular Cloud}
\author[0000-0002-3091-8061]{Anneliese M. Rilinger}
\affiliation{Department of Astronomy and Institute for Astrophysical Research, Boston University, 725 Commonwealth Avenue, Boston, MA 02215}

\author[0000-0001-9227-5949]{Catherine C. Espaillat}
\affiliation{Department of Astronomy and Institute for Astrophysical Research, Boston University, 725 Commonwealth Avenue, Boston, MA 02215}

\author[0000-0003-1283-6262]{Enrique Mac{\'\i}as}
\affiliation{Department of Astronomy and Institute for Astrophysical Research, Boston University, 725 Commonwealth Avenue, Boston, MA 02215}

\accepted{May 10, 2019 to ApJ}
\correspondingauthor{Anneliese M. Rilinger}
\email{amr5@bu.edu}

\begin{abstract}

Measuring the properties of protoplanetary disks around brown dwarfs is central to understanding the formation of brown dwarfs and their planetary companions.  We present modeling of CFHT Tau 4 and 2M0444, two brown dwarfs with protoplanetary disks in the Taurus Molecular Cloud.  By combining modeling of the spectral energy distributions and ALMA images, we obtain disk radii and masses for these objects; these parameters can be used to constrain brown dwarf formation and planet formation, respectively.  We find that the disk around CFHT Tau 4 has a total mass of 0.42 M$_{Jup}$ and a radius of 80 au; we find 2M0444's disk to have a mass of 2.05 M$_{Jup}$ and a radius of 100 au.  These radii are more consistent with those predicted by theoretical simulations of brown dwarf formation via undisturbed condensation from a mass reservoir than those predicted by ejection from the formation region. Furthermore, the disk mass of 2M0444 suggests that planet formation may be possible in this disk, although the disk of CFHT Tau 4 is likely not massive enough to form planets.  The disk properties measured here provide constraints to theoretical models of brown dwarf formation and the formation of their planetary companions.
\end{abstract}


\section{Introduction}
Since the discovery of the first brown dwarf (BD) by \citet{rebolo95}, the formation mechanism of these substellar (M $<$ 0.08 M$_{\odot}$) objects has been the subject of much debate.  Why BDs do not accrete enough mass to become fully fledged stars is still unknown.  Potential formation mechanisms of BDs can be grouped into two categories.  In the first category, hereafter ``non-ejection,'' the BD condenses from its mass reservoir without significant disruption \citep{hennebelle09, andre12, riaz18}, though some of the material may be dispersed and/or accreted by other nearby objects or the BD itself. In the second category, hereafter ``ejection,'' the BD is separated abruptly from the mass reservoir via an impulsive interaction with one or more stars \citep{reipurth01, bate03, basu12, stamatellos09}.  In the case that the BD condenses out of a disk, when ejected, the BD may already have ceased growing by accretion since the BD can only accrete the mass in a small range of angular momentum.  These two formation mechanisms are not mutually exclusive; it may be the case that some BDs form via non-ejection, and others form via ejection.
See \citet{luhman12} and \citet{whitworth18} for a complete overview of BD formation mechanisms.

BDs are known to harbor protoplanetary disks of gas and dust \citep{comeron98, natta01, muench01, natta02}.  In many ways, these disks are similar to those found around young stellar objects (e.g., T Tauri stars),  although BD disks are usually smaller \citep{hendler17}. In both cases, the disks are typically flat \citep{scholz07}, show signs of grain growth in their inner regions \citep{sterzik04, apai05, meru13, pinilla13}, and some show evidence of inner disk clearing \citep{muzerolle06}.  The disk fractions exhibited by stars are similar to the disk fractions observed for BDs, which may indicate a similar formation mechanism for the two populations \citep{luhman05}.

Notably, studying these disks can constrain the mechanism via which the BDs formed.  If a BD formed according to one of the ejection theories, its disk may have been truncated in close encounters during the ejection process and would thus have a smaller radius \citep[usually $<$ 40 au,][]{bate03, bate09, bate12}.  A larger disk radius, on the other hand, would indicate that the BD more likely formed via the non-ejection scenario without any environmental factors to truncate the disk.  Using ALMA continuum observations to measure the radii of BD disks can lend credence to one of the two methods of BD formation.

BDs have also been observed to have planetary-mass companions \citep{han13, udalski15, shvartzvald17}.  Planets are known to be common around M dwarf stars \citep{dressing13, bonfils13}, which may indicate that planets are also common around BDs.  Protoplanetary disks around BDs are thought to potentially be sites of planet formation \citep{apai05}.  Thus, constraints on planet formation around BDs can be determined by studying the properties of their protoplanetary disks.  In particular, the mass of the planet(s) that can form around the object can be determined by calculating how much mass is present in the disk.  In BD disks with a few Jupiter masses of material, enough material is present for planets of a few Earth masses to form; the mass of potential planets decreases as the mass of the disk decreases \citep{payne07}.

Here we present measurements of the radius and mass for two protoplanetary disks around BDs in the Taurus Molecular Cloud.  These objects, 2MASS J04394748+2601407 and 2MASS J04442713+2512164 (hereafter CFHT Tau 4 and 2M0444, respectively), are two of the brightest BD disk systems and have been observed at many wavelengths; hence, they are ideal targets for this study.  Both objects have been shown to have excess emission at infrared and millimeter wavelengths, indicating the presence of circum(sub)stellar disks: millimeter-wavelength dust emission \citep{klein03} and mid-infrared excess \citep{pascucci03} indicate the presence of a disk around CFHT Tau 4; excess mid-infrared \citep{guieu07} and millimeter emission \citep{scholz06} imply a disk surrounding 2M0444.  Both objects were classified as class II objects by \citet{luhman10} based on their spectral slopes.

\citet{apai04} obtained mid-infrared observations of CFHT Tau 4 and, by modeling the 10 $\mu$m silicate emission feature, determined that both grain growth and dust settling had occurred in the disk.  The spectral energy distribution (SED) of CFHT Tau 4 was fit by \citet{hendler17}, who obtained a value of $5.2\times10^{-3}$ M$_{Jup}$ for the mass of dust in the disk.  \citet{hendler17} also present a value of $78^{+43}_{-66}$ au for the disk radius, but without high-resolution interferometry observations, the uncertainties on the value are large.

The SED of 2M0444 was previously fit by \citet{bouy08}; however, without high-resolution millimeter observations, they were unable to constrain the outer radius of the disk. \citet{ricci13} presented the first high resolution ($0.16''$) millimeter-wavelength observations of 2M0444, spatially resolving thermal dust emission from a BD disk for the first time.  These observations placed a lower limit of a few tens of au on the outer radius of this disk.  Both of these studies found evidence of dust grain growth to millimeter sizes in the disk of 2M0444.

By combining SED models with high-resolution interferometry data, we are able to impose better constraints on the mass and radius of the disks of CFHT Tau 4 and 2M0444.  Section \ref{obs} describes the ALMA observations used in this work.  In Section \ref{aar}, we explain our modeling process and the results obtained.  We interpret these results in Section \ref{disc} and present a summary of our conclusions in Section \ref{sum}.

\section{Observations}\label{obs}
Archival ALMA Band 7 (0.89 mm) continuum observations of the two BDs were used in this analysis.  Observations of CFHT Tau 4 were taken in Cycle 4 on 2017 July 6 (project code: 2016.1.01511.S) using 42 antennas in the 12 m array.  Observations of 2M0444 were taken in Cycle 3 on 2016 July 24 (project code: 2015.1.00934.S) using 39 antennas in the 12 m array.   Two 2 GHz spectral windows were used to obtain the continuum emission for each object.  For CFHT Tau 4, the spectral windows were centered on 333.806 GHz and 343.911 GHz; for 2M0444, 332.013 GHz and 344.013 GHz.  The total integration time was 4.5 minutes for CFHT Tau 4 and 46 minutes for 2M0444.  Baselines ranged from 17 m to 2600 m for CFHT Tau 4 and 15 m to 1100 m for 2M0444.

The data were calibrated using the Common Astronomy Software Applications (CASA) package \citep[version 5.1.2;][]{casa}.  The quasar J0510+1800 was observed as a flux and bandpass calibrator for both objects.  This quasar was also used for phase calibration of 2M0444; another quasar, J0438+3004, was used to calibrate the phase of CFHT Tau 4.  We used the calibration scripts as provided by ALMA staff in the archive with no additional flagging required. Deconvolved images were obtained using the task CLEAN with natural weighting in CASA.  The resulting synthesized beam size for CFHT Tau 4 is $0\rlap.''24 \times 0\rlap.''12$ (position angle = $-52.47^{\circ}$), with an rms of 0.13 mJy beam$^{-1}$, giving a peak S/N of 14.  For 2M0444, the resulting beam size is $0\rlap.''28 \times 0\rlap.''19$ (position angle = $-31.02^{\circ}$), with an rms of 0.09 mJy beam$^{-1}$, producing a peak S/N of 60. 

\begin{deluxetable*}{c c c c c c c c}
\tablecaption{Object Parameters\label{tab:bdparams}}
\tablehead{
\colhead{Object} & \colhead{Spectral Type}
 & \colhead{Temperature} & \colhead{A$_{V}$} & \colhead{Luminosity} & \colhead{Mass} & \colhead{Radius} & \colhead{Distance}\\  & & \colhead{(K)} & & \colhead{(L$_{\odot}$)} & \colhead{(M$_{\odot}$)} & \colhead{(R$_{\odot}$)} & \colhead{(pc)}
}
\startdata
CFHT Tau 4 & M7$^{\pm1}$ & 2880$^{+175}_{-165}$ & 5.67$^{\pm0.89}$ & 0.175$^{+0.054}_{-0.041}$ & 0.095$^{+0.056}_{-0.012}$ & 1.68$^{+0.03}_{-0.02}$ & 147.1$^{\pm5.1}$\\
2M0444 & M7.25$^{\pm0.25}$ & 2838$^{+42}_{-128}$ & 0.0$^{+0.48}$ & 0.028$^{+0.005}_{-0.004}$ & 0.05 & 0.69$^{+0.06}_{-0.07}$ & 141.0$^{\pm2.7}$\\
\enddata
\tablecomments{Spectral types are from \citet{luhman10} for CFHT Tau 4 and \citet{luhman04} for 2M0444. All values for CFHT Tau 4 are from \citet{andrews13}. Temperature and luminosity for 2M0444 are from \citet{luhman04}; mass for 2M0444 is from \citet{ricci13}; extinction for 2M0444 is from \citet{bouy08}.  Distances are derived from Gaia DR2 parallax measurements \citep{gaia16, gaia18}.}
\end{deluxetable*}

\section{Analysis and Results}\label{aar}
In order to fully characterize the disks of CFHT Tau 4 and 2M0444, our analysis consisted of two elements: fitting models to the SEDs and modeling the ALMA visibilities.  Although described separately here for clarity, these two elements were modeled simultaneously so as to be self-consistent.  A summary of the substellar properties is shown in Table \ref{tab:bdparams}.

\subsection{Modeling Procedure}
\subsubsection{SED Fitting}\label{sed}
The disk structure for each object was modeled in order to fit the SED of each BD (Figure~\ref{fig:seds}) and constrain the disk properties, particularly the disk mass.  The SEDs were constructed using photometric data points from visible to millimeter wavelengths taken from the VizieR catalogue access tool \citep{vizier}.  See Table \ref{tab:photometry} for complete lists of the photometry points in each SED.  The SED of 2M0444 also includes a low resolution (R $\sim60$--130) spectrum from the InfraRed Spectrograph (IRS) on the Spitzer Space Telescope \citep{houck04}.

The disk structure is modeled using the D'Alessio Irradiated Accretion Disk (DIAD) radiative transfer models \citep{diad98, diad99, diad01, diad05, diad06} which have previously been used to model BD SEDs \citep[e.g.,][]{morrow08, adame11}.  The DIAD models enforce hydrostatic equilibrium to self-consistently calculate the vertical and radial structure of each disk.  The disk is also assumed to be irradiated by the central BD.  Dust in the disk is modeled in two populations: small dust grains in the atmosphere of the disk and large dust grains in the disk midplane.  Dust particle sizes are distributed according to a power law with a power of --3.5 \citep{mathis77}.  The minimum grain size for both populations is fixed at 0.005 $\mu$m; the maximum grain sizes for the two populations are free parameters in the model (a$_{max, atm}$ and a$_{max, mid}$).  The disk surface density ($\Sigma$), important for calculating disk mass, is determined by the mass accretion rate ($\dot{M}$) and disk viscosity ($\alpha$; $\Sigma \propto \dot{M} \alpha^{-1}$).  $\alpha$ is a free parameter in the DIAD model.  We fixed $\dot{M}$ at $2\times10^{-10}$ ${\msunyr}$ for 2M0444 \citep{bouy08}; no $\dot{M}$ has been measured for CFHT Tau 4, so we adopt a value of $1\times10^{-10}$ ${\msunyr}$. We set the temperature of the inner edge or ``wall'' of the disk (T$_{wall,in}$) to that of the dust destruction temperature, which we assume to be 1400 K. Other free parameters are the inner wall scale height (H$_{in}$), disk outer radius (R$_{out}$, see section \ref{alma}), inclination (i), and dust settling ($\epsilon$).  We assumed a fixed dust-to-gas mass ratio of 0.01. In order to model a pre-transitional disk, the disk structure can be modified to include an annular gap within the disk following \citet{espaillat11}.  In this case, we have both an inner wall and an outer wall.  The outer wall has a temperature of T$_{wall,out}$ and a scale height of H$_{out}$.

The photospheres are constructed using the \citet{kh95} color table and the dereddened observed J-band magnitude of the object.  Note that although both objects have spectral types of M7 or later, the latest spectral type in the \citet{kh95} color table is M6, so a spectral type of M6 was adopted for both objects when constructing their photospheres.  Although the \citet{pecaut13} color table extends to spectral type M9, the table is incomplete at infrared wavelengths beyond K-band for spectral types later than M5.  The \citet{pecaut13} spectral type M7 colors and \citet{kh95} spectral type M6 colors show good agreement within 10\% uncertainty; thus, we opted to use the more complete \citet{kh95} table.  We scale the colors to the J-band magnitude of each BD, then interpolate to obtain the emission from the photosphere at each wavelength.

Given the complexity of DIAD, running disk models is computationally expensive.  This expense renders common statistical methods of estimating best-fit model parameters and their uncertainties (e.g., Markov-Chain Monte Carlo methods, Levenberg-Marquardt $\chi^2$ minimization, etc.) unfeasible.  We ran grids of models and determined the best-fit parameters by selecting models with minimum $\chi^2$, visually inspecting the model fits, and iteratively refining the grids.  We varied the free parameters over the following ranges: 0.25--1.0 $\mu$m for a$_{max, atm}$, 500 $\mu$m--1 cm for a$_{max, mid}$, 0.001--0.1 for $\epsilon$, 75--1400 K for T$_{wall, out}$, 0.0001--0.01 for $\alpha$, 30--100 au for R$_{out}$, and $25^{\circ}$--$80^{\circ}$ for i.  Additionally, our models of resolved ALMA continuum observations (Section \ref{alma}) allowed us to further constrain the SED model parameters.  The ranges of uncertainties presented in Table \ref{tab:bestfit} should not be considered an accurate quantitative estimate of the uncertainties, but rather a qualitative approximation to the range of possible values of the parameters.

\subsubsection{ALMA Data}\label{alma}
\begin{figure*}
    \plotone{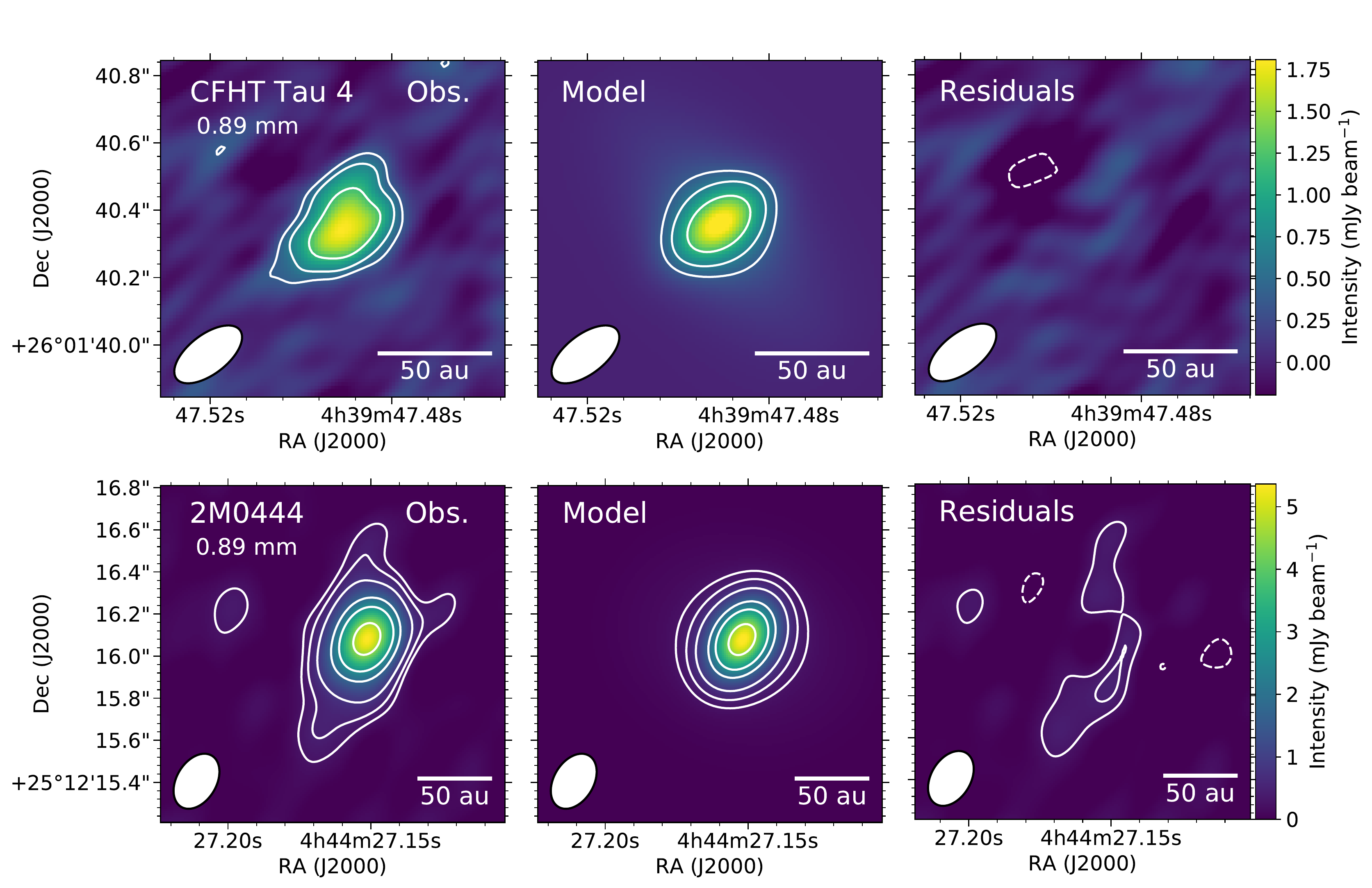}
    \caption{ALMA observations (left), models (center) and residuals (right) for CFHT Tau 4 (upper) and 2M0444 (lower). For CFHT Tau 4, the synthesized beam = $0\rlap.''24 x 0\rlap.''12$ (PA = $-52.47^{\circ}$); contour levels are --5, --3, 3, 5, 9, 20, 30, 40, 50, 60, 70, and 80 times the rms of 1.3 mJy beam$^{-1}$.
For 2M0444, the synthesized beam = $0\rlap.''28 \times 0\rlap.''19$ (PA = $-31.02^{\circ}$); contour levels are --5, --3, 3, 5, 9, 20, 30, 50, 70, 90, and 110 times the rms of 0.09 mJy beam$^{-1}$. }
    \label{fig:ALMAaplpy}
\end{figure*}

Fitting models to ALMA continuum images of protoplanetary disks can help constrain the disk radius and geometry (inclination and position angle (PA)).  Following \citet{macias18}, we modeled the ALMA Band 7 data described in Section \ref{obs}.  Using the best-fit SED models described in Section \ref{sed}, we constructed images of both objects in each spectral window.  We then used the CASA package \citep{casa} to simulate the ALMA data by Fourier transforming the synthetic model images and sampling them with the \textit{uv} coverage of the observations.  Three parameters were adjusted to improve the fit of the model image to the data: inclination, position angle, and radius.  Contours of the models and the ALMA observations were compared to assess the model fit.  See Figure \ref{fig:ALMAaplpy} for data contours, model contours, and residuals.

\renewcommand{\arraystretch}{1.0}
\startlongtable
\begin{deluxetable*}{c c c c}
\tablecaption{Photometry\label{tab:photometry}}
\tablehead{
\colhead{Wavelength ($\mu$m)} & \multicolumn{2}{c}{Flux (mJy)} & \colhead{Notes}\\
 & \colhead{CFHT Tau 4} & \colhead{2M0444} & 
}
\startdata
0.352 & ... & 0.0193 & SDSS:u\\
0.444 & ... & 0.0740* & Johnson:B\\
0.468 & ... & 0.0888* & POSS-II:J\\
0.478 & 1.59x10$^{-3}$ & 0.0945 & PAN-STARRS/PS1:g\\
0.482 & 2.57x10$^{-3}$* & 0.0534 & SDSS:g\\
0.613 & 0.0302 & 0.384 & PAN-STARRS/PS1:r\\
0.625 & 0.0410 & 0.259 & SDSS:r\\
0.640 & 0.0746* & 0.592* & POSS-II:F\\
0.749 & 0.349 & 3.79 & PAN-STARRS/PS1:i\\
0.764 & 0.544 & 2.07 & SDSS:i\\
0.784 & 1.06* & 4.44* & POSS-II:i\\
0.866 & 1.75 & 9.92 & PAN-STARRS/PS1:z\\
0.882 & 2.23 & 10.5 & UKIDSS:Z\\
0.902 & 3.44 & 8.85 & SDSS:z\\
0.960 & 4.89 & 14 & PAN-STARRS/PS1:y\\
1.031 & 8.15 & 22.6 & UKIDSS:Y\\
1.240 & 21.4 & 20.9 & 2MASS:J\\
1.249 & 20.5 & 34.8 & UKIDSS:J\\
1.251 & 21.7 & 21.4 & Johnson:J\\
1.631 & 40.6 & 29.5 & Johnson:H\\
1.651 & 41.5 & 30 & 2MASS:H\\
2.165 & 49.7 & 33.5 & 2MASS:Ks\\
2.192 & 47 & 32.5 & Johnson:K\\
2.202 & ... & 38.2 & UKIDSS:K\\
3.352 & 36.3 & 29.4 & WISE:W1\\
3.552 & 42.8 & 38.3 & Spitzer/IRAC\\
4.496 & 42.3 & 37.5 & Spitzer/IRAC\\
4.603 & 44 & 31 & WISE:W2\\
5.735 & 41.8 & 38.4 & Spitzer/IRAC\\
7.878 & 49.6 & 51.9 & Spitzer/IRAC\\
11.568 & 44 & 57.4 & WISE:W3\\
22.106 & 72.6 & 158 & WISE:W4\\
23.691 & 74.8 & 136 & Spitzer/MIPS\\
70.048 & 109 & ... & Herschel/PACS\\
71.469 & ... & 153* & Spitzer/MIPS\\
156.006 & ... & 152 & Spitzer/MIPS\\
160.111 & 150 & ... & Herschel/PACS\\
450.308 & ... & 36* & SCUBA\\
849.858 & 10.8* & ... & SCUBA-2\\
887.574 & 4.30 & 9 & ALMA Band 7\\
1300.898 & 2.38* & 5.2 & IRAM\\
1333.333 & 2.0* & 4.9 & SMA\\
3225.806 & ... & 0.87 & ALMA Band 3\\
6799.637 & ... & 0.159 & VLA Q Band\\
9099.181 & ... & 0.071 & VLA Ka Band\\
13599.275 & ... & 0.064 &	VLA K Band\\
\enddata
\tablecomments{Uncertainties in flux measurements are typically below 10\%.  Fluxes denoted by an asterisk (*) have uncertainties between 10\% and 40\%.}
\end{deluxetable*}

Comparing contours by eye allowed us to broadly constrain the inclination, position angle, and radius of the disks, but radial profiles provided a more rigorous test of the model fits.  These profiles are constructed by finding the average intensity in elliptical rings in the object.

\begin{figure}
    \plotone{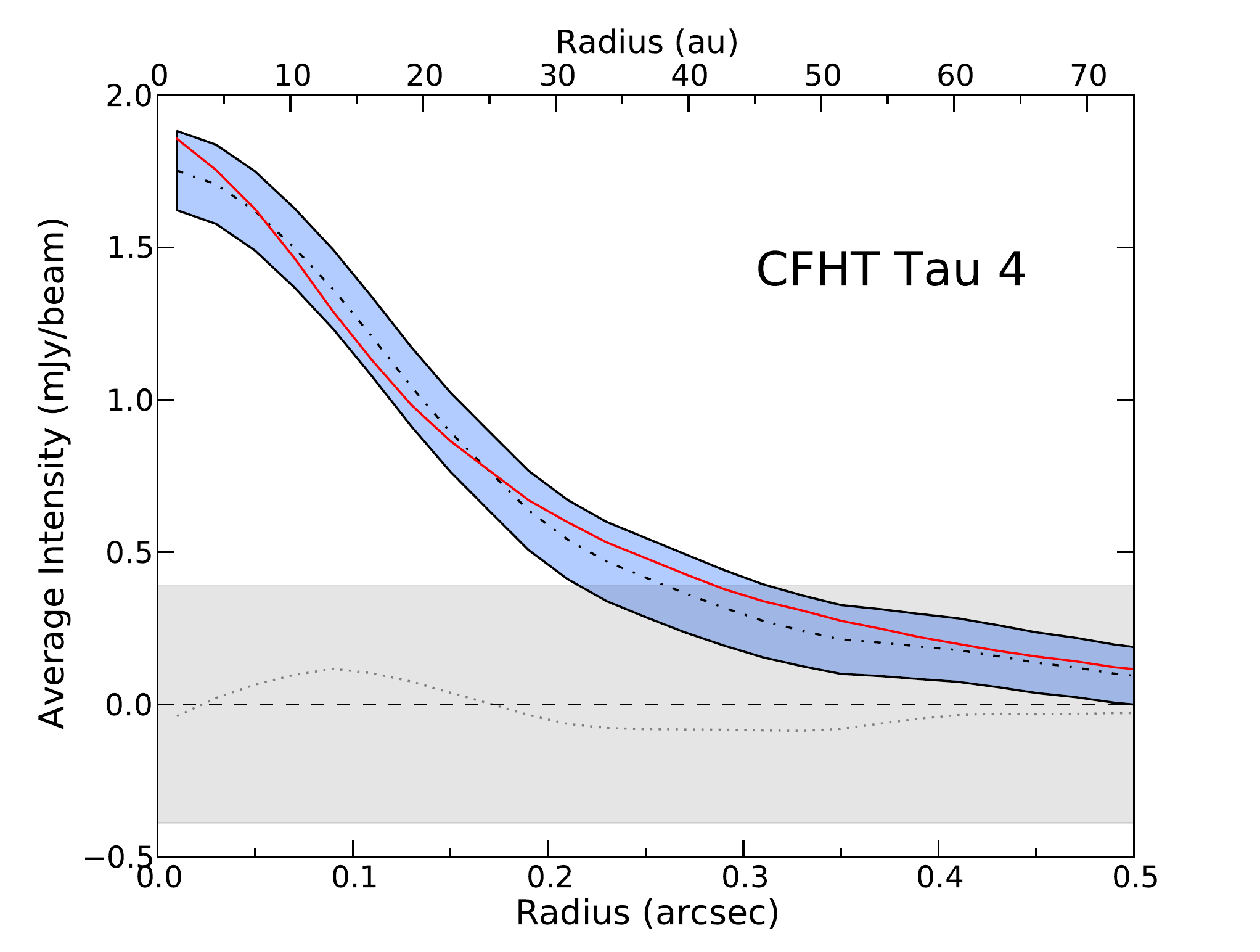}
    \caption{Radial profiles for CFHT Tau 4.  The observed profile is shown by the dot-dashed black line with a 1$\sigma$ uncertainty shown by the blue shaded region.  The model profile is shown by the red line.  Residuals are shown by the gray dotted line, with a 3$\sigma$ uncertainty shown by the gray shaded region and 0 denoted by the dashed line.}\label{fig:radialprofs}
\end{figure}

\begin{figure*}
\gridline{\fig{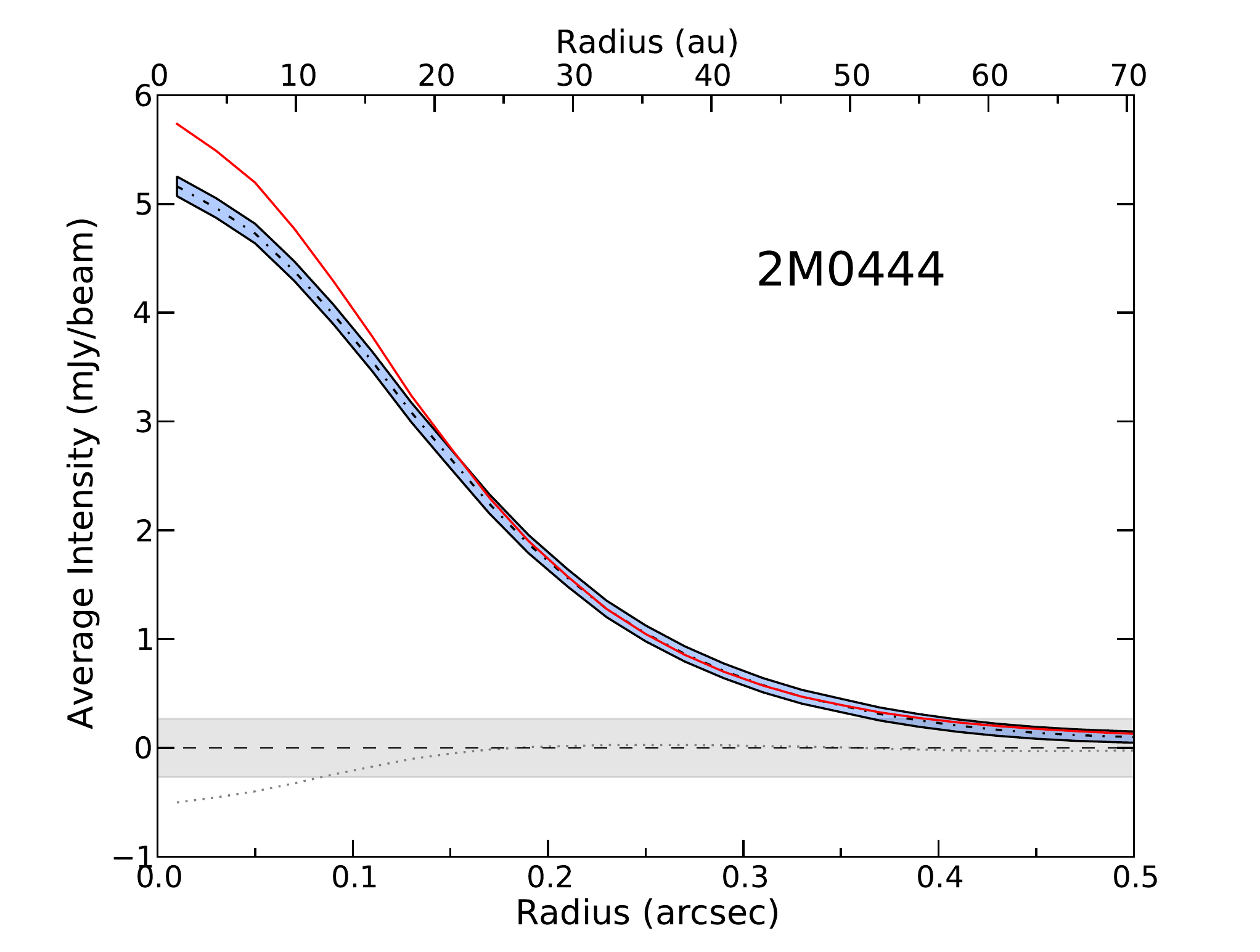}{0.5\textwidth}{}
          \fig{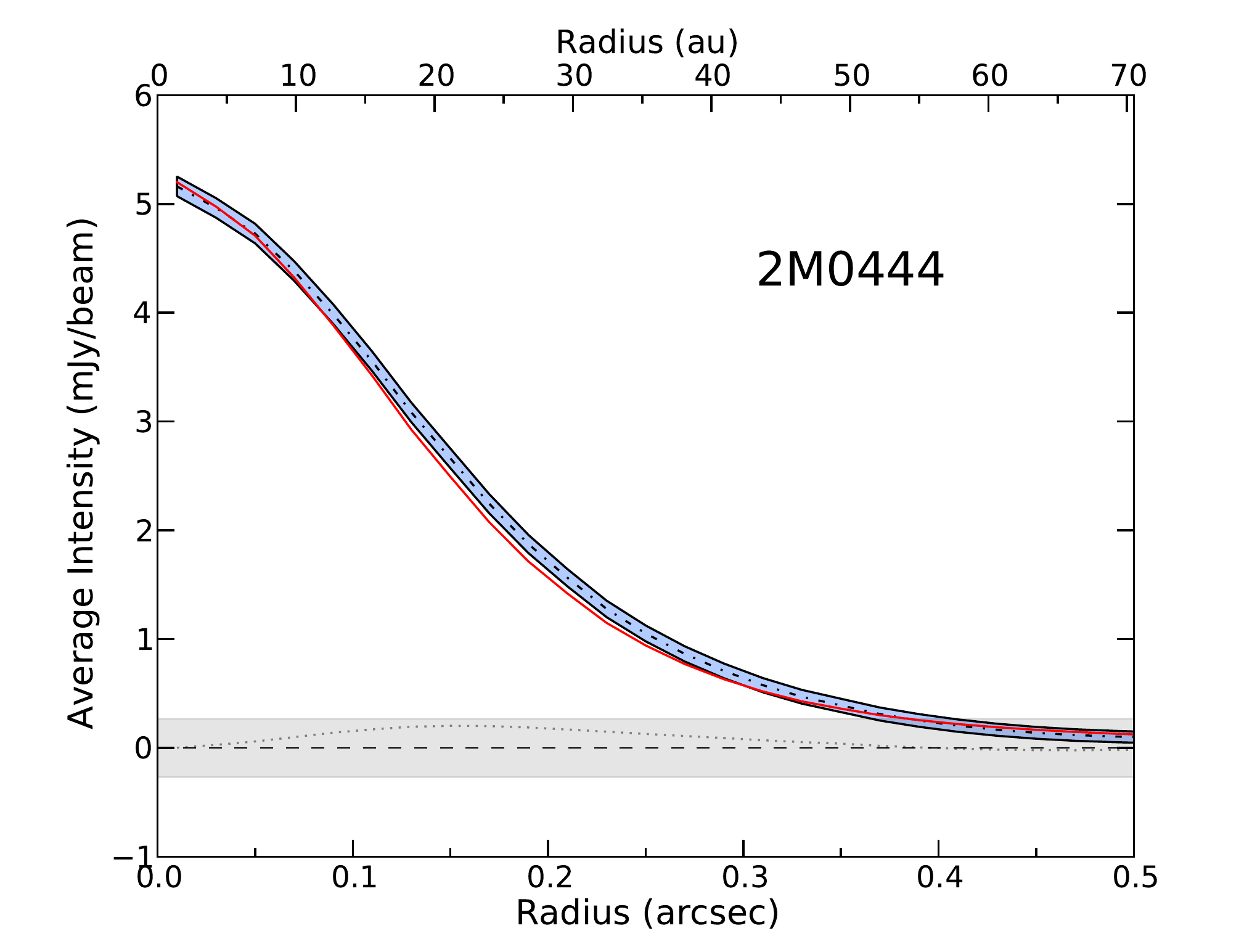}{0.5\textwidth}{}}
          \vspace*{-8mm}
        \caption{Radial profiles for 2M0444 with a full disk model (left) and a pre-transitional disk model (right). The observed profile is shown by the dashed black line with a 1$\sigma$ uncertainty shown by the blue shaded region.  The model profile is shown by the red line.  Residuals are shown by the gray dotted line, with a 3$\sigma$ uncertainty shown by the gray shaded region.}
\label{fig:ptd}
\end{figure*}

We constructed radial profiles for both the observed ALMA data and for our models.  The observed and best-fit model radial profiles for CFHT Tau 4 are shown in Figure \ref{fig:radialprofs} and the best-fit model radial profile for 2M0444 is shown in the right panel of Figure \ref{fig:ptd}.  We varied disk radius, position angle, and inclination for each object until the model profile was within 1$\sigma$ of the observed profile and the residuals were minimized.

As a final test of our models, we compared observed ALMA visibilities to our simulated model visibilities.  Visibilities are complex numbers, so we plot the real part of the visibilities against the deprojected distance in the \textit{uv} plane.  Note that larger \textit{uv} distances correspond to longer baselines and thus finer angular resolution.

\begin{figure*}
    \plotone{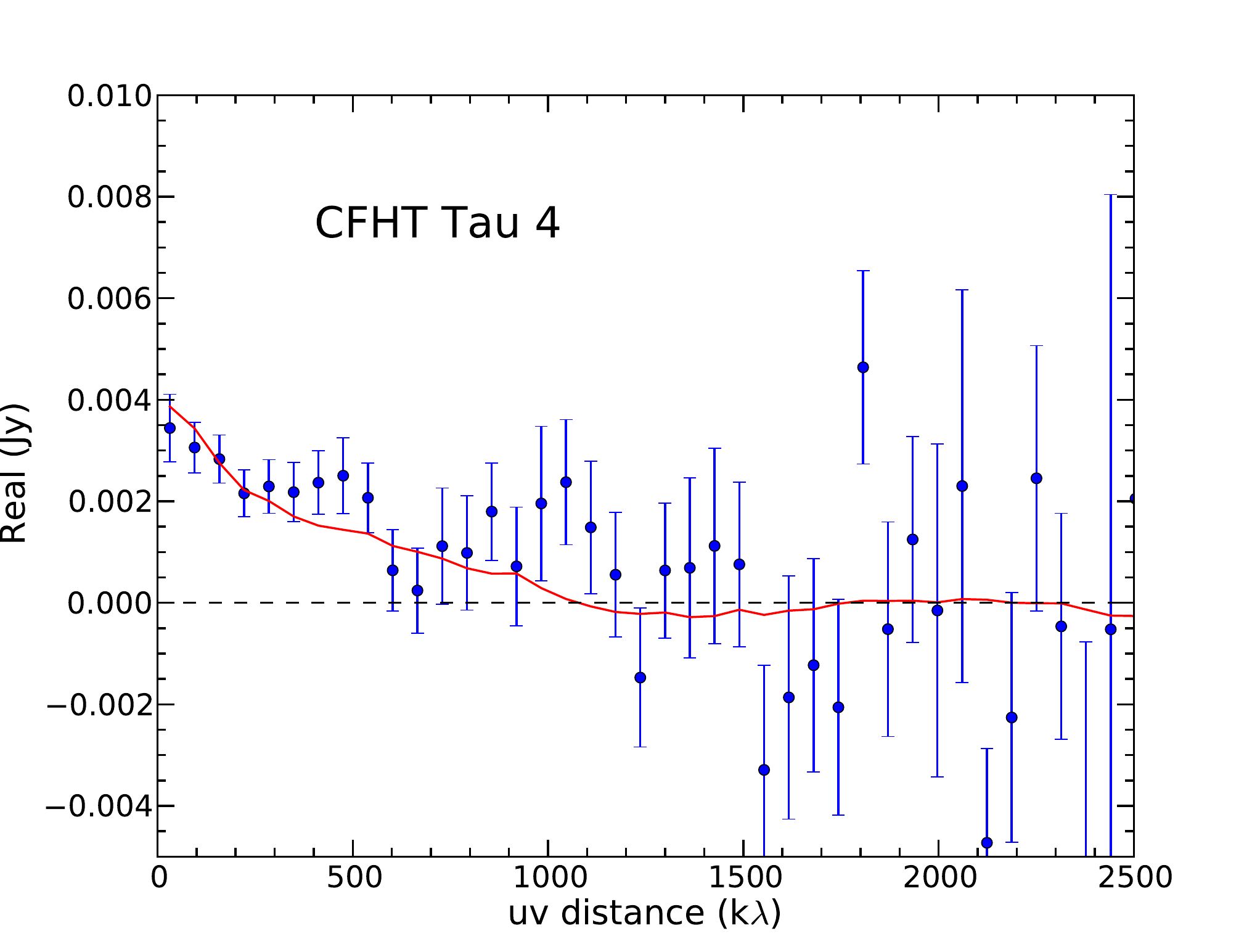}
    \caption{Observed visibilities of CFHT Tau 4 (blue points) compared to modeled visibilities (red line).}
    \label{fig:cfht_vis}
\end{figure*}

\begin{figure}
    \gridline{\fig{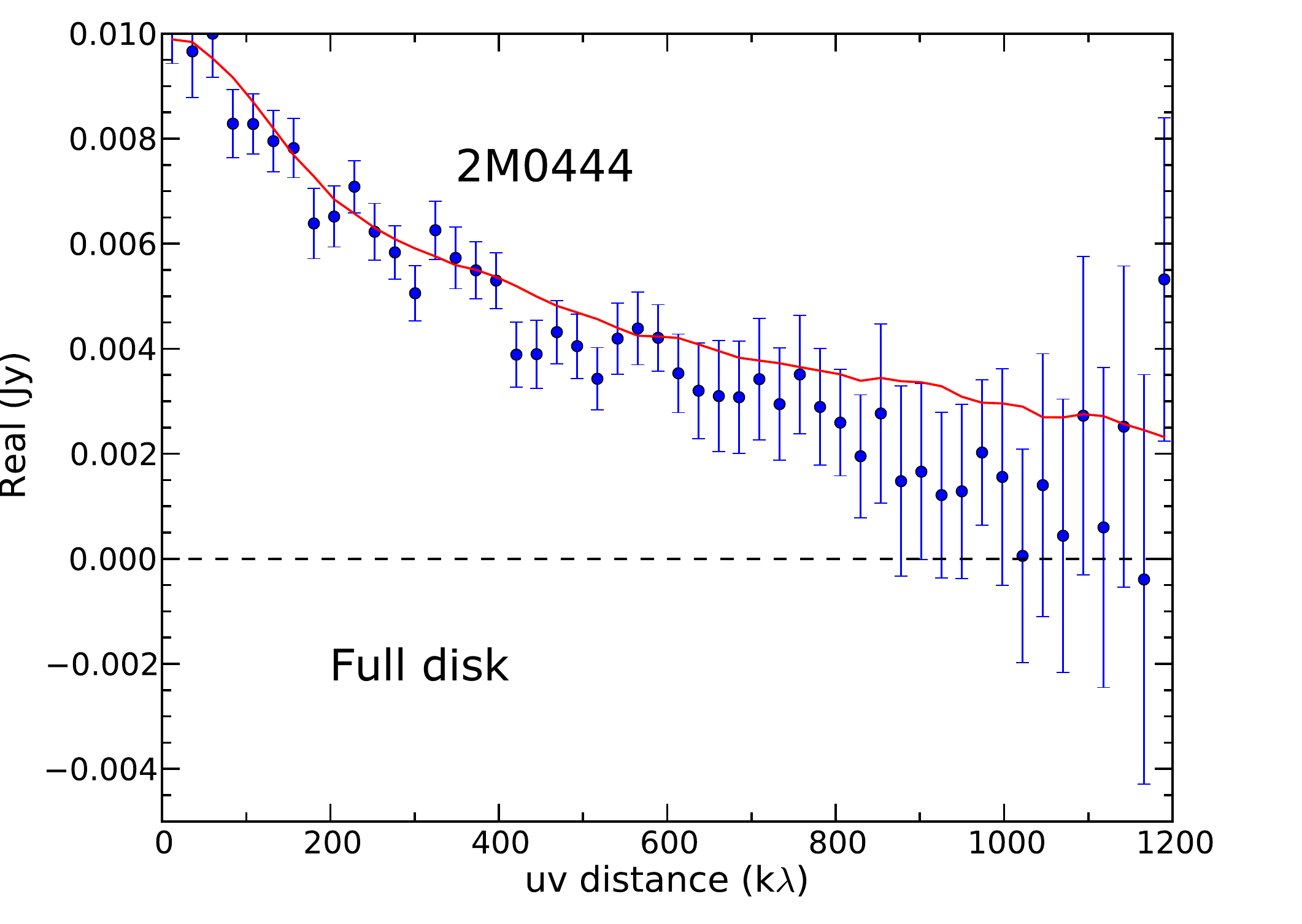}{0.5\textwidth}{}}
    \vspace*{-8mm}
    \gridline{\fig{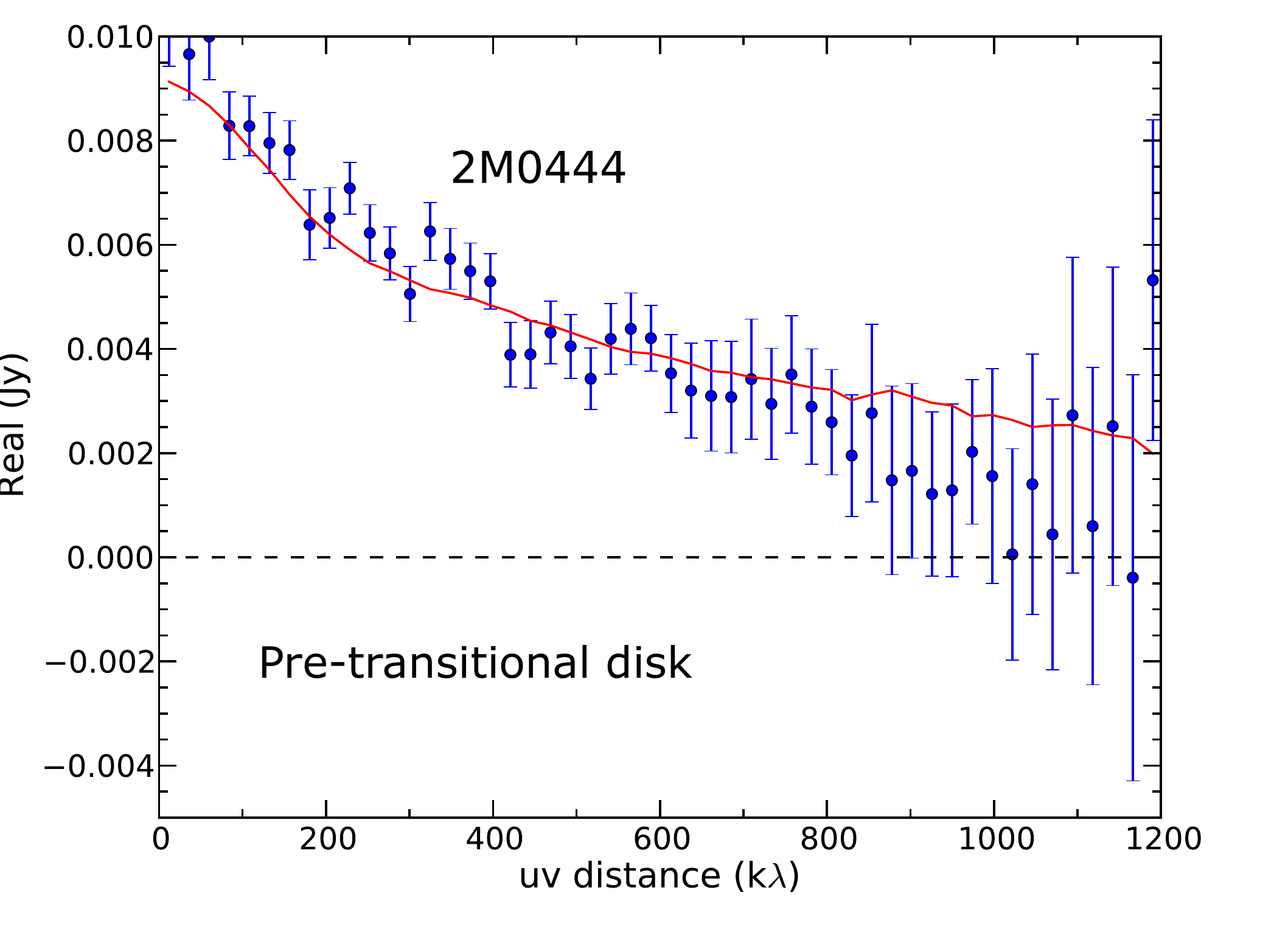}{0.5\textwidth}{}}
    \vspace*{-8mm}
    \caption{Observed visibilities of 2M0444 compared to modeled visibilities for a full-disk model (top) and a pre-transitional disk model (bottom).  Red lines indicate modeled visibilities and blue points show observed visibilities.}
    \label{fig:2M0444_vis}
\end{figure}

Plots of the observed and modeled visibilities are shown in Figures \ref{fig:cfht_vis} (for CFHT Tau 4) and \ref{fig:2M0444_vis} (for 2M0444).  In Figure \ref{fig:2M0444_vis}, we present the modeled visibilities for 2M0444 with a full disk and a pre-transitional disk.  This side-by-side comparison confirms that a pre-transitional disk model reproduces the observed visibilities of 2M0444 much better than our full-disk model.

\subsection{Model Results}
\begin{figure*}
    \gridline{\fig{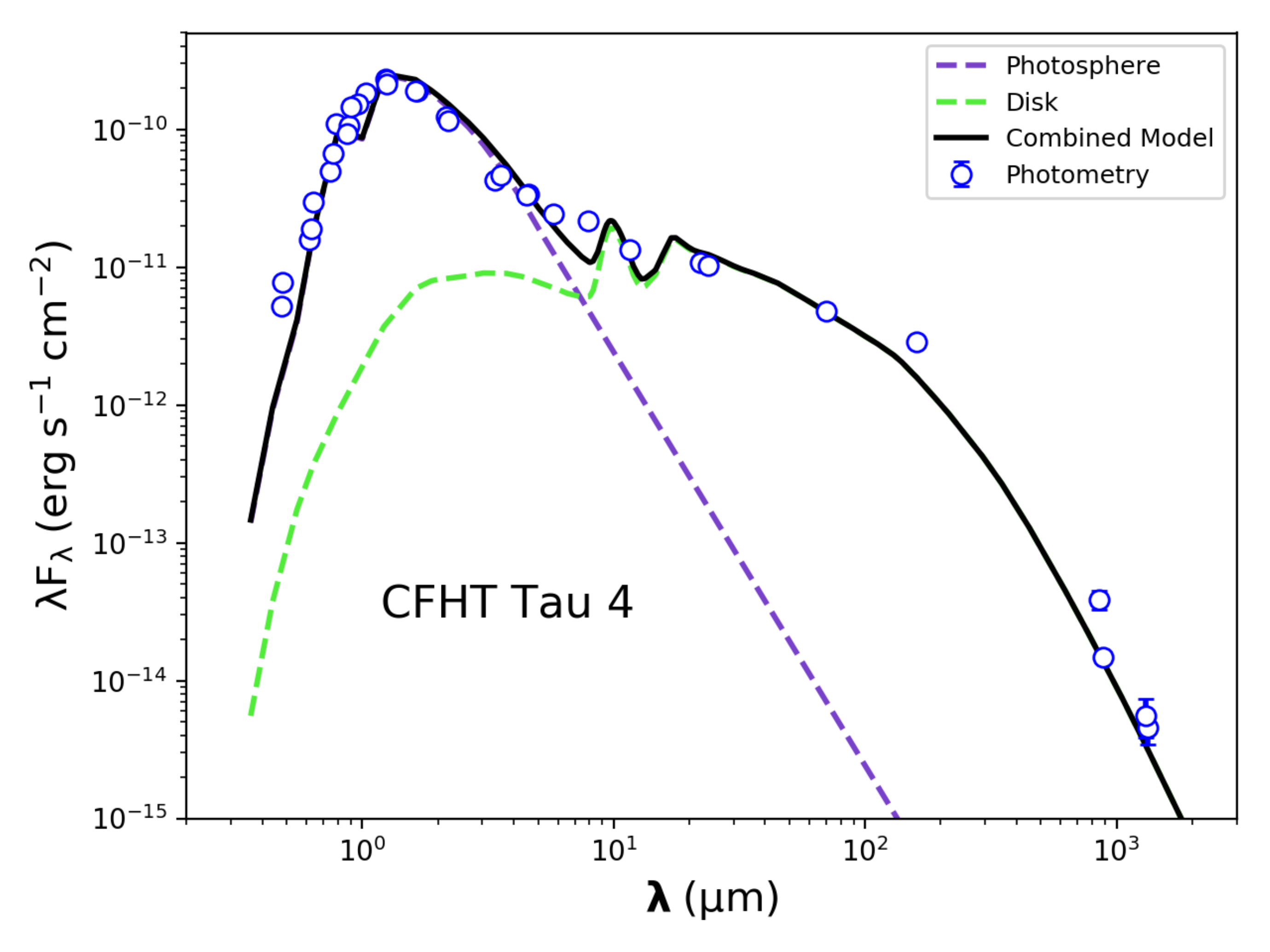}{0.5\textwidth}{}
          \fig{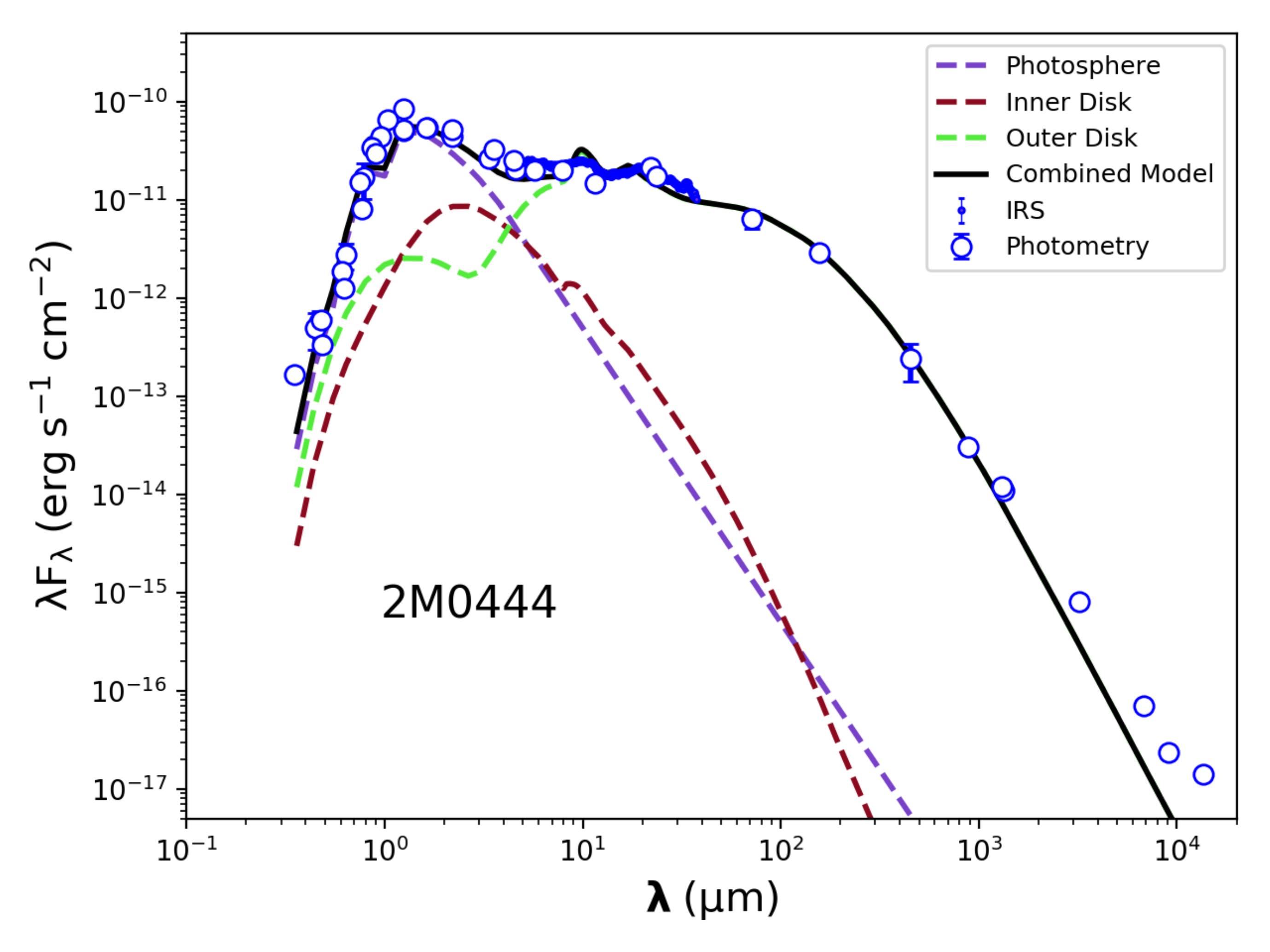}{0.5\textwidth}{}}
    \vspace*{-8mm}
    \caption{Best-fit SED models for CFHT Tau 4 (left) and 2M0444 (right).   CFHT Tau 4 is modeled as a full disk and 2M0444 is modeled as a pre-transitional disk.  Photometry is shown by open circles.  For each object, the total model is shown by a solid black line, the adopted photosphere is shown by a blue dashed line and the disk emission is shown in green.  For 2M0444, the inner disk component is shown in red.  Both SEDs are dereddened (A$_V$ given in Table \ref{tab:bdparams}).}\label{fig:seds}
\end{figure*}

The SED of CFHT Tau 4 is well fit by our full-disk model (Figure \ref{fig:seds}).  The BD dominates the SED in the optical through NIR wavelengths until $\sim$5 microns where excess emission above the photosphere due to the disk becomes apparent.  The parameters of the best-fitting model are shown in Table \ref{tab:bestfit}.  The total disk mass was computed by integrating over the surface mass density of the disk (as determined by $\dot{M}$ and the best-fit $\alpha$) over the radius of the disk.  The total mass of the disk of CFHT Tau 4 was calculated to be 0.42 M$_{Jup}$ (133 M$_{\oplus}$), assuming a dust-to-gas mass ratio of 0.01.

The SED of 2M0444 shows significant excess emission above the photosphere, particularly at mid-IR wavelengths, that could not be reproduced by a full-disk model. This mid-IR excess is consistent with that expected from a pre-transitional disk \citep{espaillat07}. Thus, we model 2M0444 as a pre-transitional disk by including a gap in the disk. In Figure \ref{fig:seds}, we show the pre-transitional disk model fit, separately plotting the emission from the inner and outer disk. The outer wall is illuminated by the BD and emits brightly in the mid-IR.  We note that the $\sim$1 micron emission from the outer disk is dominated by scattered light. Best-fit model parameters are presented in Table \ref{tab:bestfit}.  We also found that increasing the luminosity of 2M0444 within the reported uncertainties improved the overall SED fit, so we adopt a luminosity value of 0.033 L$_{\odot}$.  The total disk mass based on these model parameters was calculated to be 2.05 M$_{Jup}$ (652 M$_{\oplus}$).

We find that the best-fit radial profile model for CFHT Tau 4 has a disk radius of 80 au, a position angle of 40$^{\circ}$, and an inclination of 70$^{\circ}$.  This inclination agrees reasonably well with \citet{hendler17} and \citet{ricci14}, who both find CFHT Tau 4 to have an inclination of $75^{\circ}$--$80^{\circ}$.  The disk radius also agrees well with previous results; see Section \ref{disc:rad} for a complete discussion.

\renewcommand{\arraystretch}{1.2}
\begin{deluxetable}{c c c}
\tablecaption{Best-Fit Parameters\label{tab:bestfit}}
\tablehead{
\colhead{Parameter} & \colhead{CFHT Tau 4} & \colhead{2M0444}
}
\startdata
a$_{max, atm}$ ($\mu$m) & 0.25$^{+0.25}_{-0}$ & 1.0$\pm$0.25\\
a$_{max, mid}$ ($\mu$m) & 3000$\pm$2000 & 5000$\pm$2000\\
Dust to gas mass ratio\tablenotemark{a} & 0.01 & 0.01\\
$\epsilon$ & 0.007$^{+0.003}_{-0.002}$ & 0.05$_{-0.04}^{+0}$\\
H$_{in}$ & 1.0$^{+1}_{-0}$ & 4.0$\pm$1\\
H$_{out}$ & ... & 10.0$\pm$1\\
T$_{wall,in}$\tablenotemark{a} (K) & 1400 & 1400\\
T$_{wall,out}$ (K) & ... & 400$\pm$50\\
R$_{wall, in}$\tablenotemark{b} (au) & 0.07 & 0.07\\
R$_{wall, out}$\tablenotemark{b} (au) & ... & 0.27$\pm$0.1\\
R$_{out}$ (au) & 80$\pm$10 & 100$\pm$10\\
$\alpha$ & 0.0007$\pm$0.0001 & 0.00046$\pm$0.00001\\
$\dot{M}$\tablenotemark{c} (M$_{\odot}$ yr$^{-1}$) & 1x10$^{-10}$ & 2x10$^{-10}$\\
M$_{disk}$\tablenotemark{d} (M$_{Jup}$) & 0.42$^{+0.08}_{-0.05}$ & 2.05$^{+0.07}_{-0.06}$\\
i ($^{\circ}$) & 70$\pm$10 & 40$\pm$10\\
PA ($^{\circ}$) & 40$\pm$10 & 70$\pm$10\\
\enddata
\tablenotetext{a}{Fixed to typically assumed values.}
\tablenotetext{b}{Calculated using T$_{wall}$ following \citet{diad05}.}
\tablenotetext{c}{Fixed.  We adopt $\dot{M}$ for 2M0444 from \citet{bouy08} and assume a low $\dot{M}$ for CFHT Tau 4 (see Section \ref{sed}).}
\tablenotetext{d}{Calculated using $\dot{M}$ and $\alpha$ following \citet{diad05}}
\end{deluxetable}

We find that the best-fit radial profile for 2M0444 has a disk radius of 100 au, a position angle of 70$^{\circ}$, and an inclination of 40$^{\circ}$.  This inclination is consistent with \citet{bouy08}, who claim it should be less than 55$^{\circ}$. A complete discussion of the disk radius is given in Section \ref{disc:rad}.

\section{Discussion}\label{disc}
\subsection{Disk Radius and BD Formation}\label{disc:rad}
The disk radii we obtain for CFHT Tau 4 and 2M0444 agree well with previous studies. Using radiative transfer models to fit the SED of CFHT Tau 4, \citet{hendler17} obtained a disk radius of $78^{+43}_{-66}$ au.  \citet{ricci14} used $0\rlap.''47$ resolution continuum ALMA observations to place a lower limit of 80 au on the disk radius of CFHT Tau 4.  Our value of 80 au agrees with both of these results.  The disk radius of 2M0444 was constrained by \citet{bouy08} and \citet{ricci13}, both of whom determined via SED modeling that the radius of the disk must be greater than 10 au.  \citet{bouy08} adopted a value of 300 au for the disk radius, but they note that disk radius is not well constrained with submillimeter/millimeter photometry alone.  \citet{ricci14} obtain a value of $139^{+20}_{-27}$ au for the disk radius of 2M0444 by modeling ALMA data.  We find the radius to be somewhat smaller at 100~au but still in good agreement with their lower limits. 

The disk radii obtained from this analysis point toward a non-ejection formation mechanism for these BDs, consistent with previous studies by \citet{ricci13, ricci14}.  In the ejection scenarios suggested by \citet{reipurth01} and \citet{bate03}, protoplanetary disks around BDs would be truncated to smaller radii via interactions between the disk and BD as well as interactions within the disk itself.  \citet{bate12} showed that in this scenario, only 20\% of disks around BDs would have radii greater than 40 au and $\lesssim$ 10\% of disks would have radii greater than 100 au.  Given the large radii of the disks presented here, it is unlikely that these objects formed via ejection.  However, once an ejected BD becomes ``free-floating'', its remaining disk is expected to spread, and in doing so be accreted onto the BD on a time-scale determined by the effective viscosity in the disk \citep{lyndenbell74, pringle81}.  Over this time, the mass of the disk will decrease by about an order of magnitude, so its surface density will go down by a factor of $\sim$1000, and the IR and mm emission will decrease. The selection effects resulting from the evolution of the disk, and the timescale on which this evolution occurs, make firm conclusions difficult.

The disks studied here are among the brightest protoplanetary disks around BDs.  It is possible that these disks may also be among the largest, and thus other, fainter, disks may be smaller and more likely to have formed via ejection.  \citet{testi16} observed 17 BDs and very-low-mass stars in $\rho$ Ophiuchus, of which 11 were found to have disks; of these 11 disks, only 2 were resolved well enough to measure the disk radii.  The resolved disks were found to have radii of $\sim20$ au, and the radii of the unresolved disks may be even smaller.  \citet{testi16} claim that most BD disks may fall into this smaller radius range and only the brightest disks extend to larger radii, which follows the distribution presented by \citet{bate12}.  Furthermore, \citet{testi16} suggest that the larger radii of CFHT Tau 4 and 2M0444 may be accounted for by the older age and lower density of the Taurus Molecular Cloud.  More high-resolution ALMA data of faint BD disks will allow us to measure the radii of these disks to determine whether CFHT Tau 4 and 2M0444 are outliers.

\subsection{Disk Mass and Planet Formation}
Using the DIAD model, as described in Section \ref{sed}, we calculate the total disk masses of CFHT Tau 4 and 2M0444 to be 0.42 M$_{Jup}$ (133 M$_{\oplus}$) and 2.05 M$_{Jup}$ (652 M$_{\oplus}$), respectively.  Given uncertainties in the dust-to-gas mass ratio, we compare the dust masses we find to previously reported values.  We find the dust mass of CFHT Tau 4 to be 0.0042 M$_{Jup}$ (1.33 M$_{\oplus}$) and the dust mass of 2M0444 to be 0.0205 M$_{Jup}$ (6.52 M$_{\oplus}$).  \citet{klein03} found the dust mass of CFHT Tau 4 to be 1.4--7.6 M$_{\oplus}$, consistent with our value.  From their 1.3 mm flux measurement, \citet{scholz06} find the dust disk mass of CFHT Tau 4 to be $0.0071\pm0.0022$ M$_{Jup}$.  We find the dust mass of CFHT Tau 4 to be 0.0042 M$_{Jup}$, which is slightly lower but still in reasonable agreement with the \citet{scholz06} value.  Our value for the mass of 2M0444 is also consistent with previous measurements.  From their SED fit for 2M0444, \citet{bouy08} find the disk of 2M0444 to have a dust mass on the order of 0.01 M$_{Jup}$; \citet{scholz06} obtained a dust mass of $0.0225\pm0.0027$ M$_{Jup}$ based on their 1.3 mm flux observation.

We can assess whether planet formation is likely to occur in these BD disks by comparing our results to the core accretion models in \citet{payne07}.  Their simulations form Earth-mass planets frequently around BDs with disks of a few Jupiter masses of material; however, the simulations very rarely produced planets in BD disks with only a fraction of the mass of Jupiter.  \citet{payne07} suggest that the lower disk surface density in the less massive disks accounts for the deficit of planets.  Based on this result, the disk of 2M0444 contains just enough material for Earth-mass planet formation to occur.  However, CFHT Tau 4 is unlikely to form Earth-mass planets, given that its total mass is less than 1 M$_{Jup}$.  Neither object contains enough material for Jupiter-mass planets to form, which could suggest that observed gas giant companions to BDs \citep[e.g.,][]{chauvin04, joergens07, todorov10, han13} do not form in protoplanetary disks and instead must form via some other mechanism.  Many of these observed Jupiter-mass companion systems have large mass ratios and wide separations, which may indicate that such systems form via gravitational fragmentation of massive primordial disks \citep{lodato05}.

Previous work has found other protoplanetary disks to be similar to CFHT Tau 4 in that they are not massive enough to form planets.  For example, \citet{testi16} find that most of the BD disks in their sample contain only $\sim1$ M$_{\oplus}$ of dust.  This result is also consistent with other studies, e.g., \citet{najita14}, who found that the masses of protoplanetary disks around T Tauri stars are too small to explain known planetary systems.  Given these findings and its large radius, 2M0444 may be an outlier in its ability to form planets.  The formation of observed planets around BDs and very-low-mass stars \citep[e.g., the TRAPPIST-1 system;][]{gillon16, gillon17} thus remains mysterious: If only a small fraction of protoplanetary disks are massive enough to form planets, why are planetary systems commonly observed?  

Our results could still be consistent with the presence of planetary systems around BDs if a planet has already formed in the disk.  Rapid planet formation has also been proposed by \citet{manara18}, who suggest planet formation within $<$0.1--1 Myr.  Evidence of dust processing and settling in young ($\sim1$ Myr old) disks (e.g., \citet{grant18} and references therein) also supports planet formation on timescales of $<$1 Myr.  Furthermore, planet formation on these timescales is consistent with observations of ring substructures in young protoplanetary disks, such as in HL Tau \citep[$<1$ Myr old,][]{alma15}.  Similarly, a significant fraction of the dust mass may be in bodies smaller than planets but larger than $\sim1$ m, which do not contribute to the dust emissivity at millimeter wavelengths.

Alternatively, the disks may be replenished by gas and small dust grains from the surrounding interstellar medium throughout their lifetimes such that the overall amount of planet-forming material is greater than that present in the disk at any one time \citep[][and references therein]{manara18}. Observations of Class 0/I objects will help assess these theories.

It is important to note that the mass discrepancy may simply be due to uncertainties in mass determination \citep{bergin18}.  However, \citet{manara18} argue that the uncertainty in disk masses would need to be systematically underestimated by an order of magnitude or more, which would conflict with other observations.  Specifically, higher disk mass would conflict with the observed agreement between current disk mass estimates and mass accretion rates with theoretical expectations \citep{manara16, rosotti17, lodato17}.  Furthermore, the observed faint CO lines in disks would be difficult to explain with increased disk mass \citep{miotello17, long17}. 

\subsection{Gap in the Disk of 2M0444}\label{cavity}
We find that the disk of 2M0444 is best fit with a pre-transitional disk model with a gap in the disk from 0.02 au to 0.27 au.  Transitional disks (i.e., disks with holes with large depletions of dust in the inner disk) have been previously inferred around one BD based on its SED \citep[2MASS J03445771+3207416;][]{muzerolle06} and imaged around a very-low-mass star \citep[CIDA 1;][]{pinilla18}; 2M0444 would be the first inferred pre-transitional BD disk.  Given that pre-transitional and transitional disks are not uncommon around around T Tauri stars \citep[e.g.,][]{espaillat07, muzerolle10, kim13}, BD disks may exhibit the same types of structures.

Gaps in protoplanetary disks have been proposed to be due to several mechanisms, including snowlines, photoevaporation, and planet formation.
The small gap in 2M0444 may be explained by higher dust grain fragmentation at a snowline (e.g., water, CO), which could result in a lack of large dust particles within the snowline \citep{zhang15}.  Outside a snowline, the molecule freezes onto dust particles, increasing their size \citep{gundlach15}.  These large particles dynamically decouple from the gas in the disk and can migrate inside the snowline, where the molecule sublimates; the dust particles are more prone to fragmentation and thus they decrease in size.  Therefore, a lack of large dust grains is expected within a snowline \citep{pinilla17}.
To check if a snowline is creating the gap in 2M0444, we constructed a temperature profile of the disk and compared the temperature at 0.27 au to the condensation temperatures of various molecules presented in \citet{zhang15}.  We find the temperature of the disk at 0.27 au to be 117 K.  The snowlines of common molecules were either too warm (H$_2$O at 128--150 K) or too cold (CO at 23--28 K, CO$_2$ at 60--72 K) to explain the gap at 0.27 au.

Another proposed explanation of gaps in some disks is photoevaporation \citep[e.g.,][]{clarke01, alexander06, owen10}.  According to this mechanism, a small gap should open in the outer disk, and the inner disk should accrete onto the star. Given the low accretion rate of 2M0444, its clearing could potentially be due to photoevaporation \citep{owen12}.  However, photoevaporation models more easily explain disks with holes and not gaps, given that the isolated inner disk should accrete quickly onto the star once the gap has opened.

A third explanation for the presence of a gap in the disk of 2M0444 is a forming planet.  As planets form in a protoplanetary disk, they clear material from the disk \citep{kley12, baruteau14}.  Planets are able to reproduce the substructure seen in several disks \citep{zhang18}.  A planet forming close to the central BD in 2M0444 could potentially clear out the material in the innermost 0.27 au as it forms in the disk.  

SED modeling is sensitive to small gaps in the inner disk, but not to those in the outer disk \citep{espaillat10}. ALMA will not be able to resolve the small inner disk gap in 2M0444 that is inferred here. However, ALMA can resolve small gaps further out in the disk to which our SED modeling was not sensitive. High-resolution images of BD disks are necessary to determine whether they contain similar substructures as those seen in disks around more massive stars \citep{dsharp}, and if so, whether these substructures are created by the presence of forming planets.

\section{Summary}\label{sum}
We obtained measurements of the disk mass and radius for 2MASS J04394748+2601407 (CFHT Tau 4) and 2MASS J04442713+2512164 (2M0444), two protoplanetary disks around BDs in the Taurus Molecular Cloud, by simultaneously fitting their SED and ALMA Band 7 continuum data in order to assess proposed formation mechanisms for the BDs as well as the potential for planet formation in the disks.

We find that the disk of 2M0444 contains enough material to form Earth-mass planets (2.05 M$_{Jup}$), but the disk of CFHT Tau 4 (0.42 M$_{Jup}$) is not massive enough for planets to form, based on the results from \citet{payne07}.  Other works find disks around BDs with similar masses to that of CFHT Tau 4; 02M0444 may be unusual in its ability to form planets.

The disk radii we determine from our ALMA models (80 au for CFHT Tau 4, 100 au for 2M0444) suggest that both objects likely did not form via ejection from their formation region; these radii are larger than those expected from the \citet{bate03} and \citet{bate12} ejection simulations.  However, given that the disks studied here are some of the brightest known BD disks, these disks may be among the largest BD disks; other, fainter BDs may be surrounded by smaller disks.  More ALMA observations of dimmer BD protoplanetary disks are necessary to better understand the overall distribution of disk radii, which will in turn inform us about the formation mechanism of these objects.

Finally, our models indicate that the disk of 2M0444 is a pre-transitional disk with a gap between 0.02 au and 0.27 au.  High-resolution observations with future facilities such as the ngVLA may be able to resolve this gap.

\vspace{\baselineskip}
This paper utilizes the D'Alessio Irradiated Accretion Disk (DIAD) code.  We wish to recognize the work of Paola D'Alessio, who passed away in 2013. Her legacy and pioneering work live on through her substantial contributions to the field.  We thank the referee for their constructive comments.  We thank Sarah Luettgen and Connor Robinson for helpful discussions.  The authors acknowledge support from the National Science Foundation under Career grant AST-1455042 and the Sloan Foundation.

This paper makes use of the following ALMA data: ADS/JAO.ALMA \#2016.1.01511.S and \#2015.1.00934.S. ALMA is a partnership of ESO (representing its member states), NSF (USA) and NINS (Japan), together with NRC (Canada), MOST and ASIAA (Taiwan), and KASI (Republic of Korea), in cooperation with the Republic of Chile. The Joint ALMA Observatory is operated by ESO, AUI/NRAO and NAOJ.  The National Radio Astronomy Observatory is a facility of the National Science Foundation operated under cooperative agreement by Associated Universities, Inc.

\software{CASA (v5.1.2; McMullin et al. 2007)}

\end{document}